\documentstyle[12pt]{article}
\begin{document}
\medskip

\pagestyle{empty}
             \rightline{
                            MRI-PHY/94/17
             }

              \vspace{5 mm}

                    \begin{center}

                         {\bf LOCAL SCALING OF TIME IN HAMILTONIAN PATH
                         INTEGRATION }

                     \vspace{1.5 cm}

                                $  {\rm   A.K. Kapoor }^\dagger $ \\
                                          Mehta Research Institute, \\
                               10 Kasturba Gandhi Road Allahabad 211002
                               INDIA \\
                     \vspace{3 mm}
                                                  and

                     \vspace{3 mm}

                                                Pankaj Sharan \\
                                  Physics Department, Jamia Millia Islamia,
                                  \\
                                    Jamia Nagar, New Delhi 110025, INDIA \\

                     \vspace{0.5 cm}

                                                {\bf ABSTRACT }

                                                       \

                     \end{center}
Inspired  by  the  usefulness  of local scaling of time in the path integral
formalism,  we  introduce  a  new  kind of hamiltonian path integral in this
paper.  A  special  case  of  this  new  type  of  path  integral  has  been
earlier found  useful  in  formulating  a scheme of hamiltonian path integral
quantization  in  arbitrary  coordinates. This scheme has the unique feature
that   quantization  in  arbitrary co-ordinates  requires  hamiltonian  path
integral  to  be  set up in terms of the classical hamiltonian only, without
addition of any adhoc $ O(\hbar ^2) $terms. In this paper  we  further study
the  properties of hamiltonian path integrals in arbitrary co-ordinates with
and  without  local  scaling  of  time  and  obtain the Schrodinger equation
implied by the hamiltonian path integrals. As a simple  illustrative example
of  quantization  in  arbitrary coordinates and of exact path integration we
apply   the  results  obtained  to  the  case  of  Coulomb  problem  in  two
dimensions.

\vspace{0.5 cm}
\hrule
                \noindent
$^\dagger $ Permanent Address : University of Hyderabad, Hyderabad 500134,
INDIA   \\
\ \  email : ashok@mri.ernet.in \\

\pagebreak
\pagestyle{plain}
\noindent
    {\bf 1. Introduction}
\vspace{5 mm}

 The  method of path integration $^1$  is  an  attractive  way  of obtaining
 the  quantum mechanical amplitudes. There  are  two main approaches to path
 integrals. One of  them  is  to  define them rigorously in terms of a class
 of  paths  and  measures  defined  on  the  class  of paths $^2.$ The other
 approach is more widely practiced. For reviews of the path integral  method
 and its applications we refer the reader to references 3-9. The  so  called
 'time   slicing'  method  contains,  essentially,  nothing  more  than  the
 observation  that  the  propagator,  being  the  matrix element of $ \exp[-
 itH/\hbar ] $ ,can  be written as an $ N \rightarrow \infty $ limit of an $
 N-$  fold  interal  of   the   matrix  elements of $ \exp[-itH/\hbar].$ The
 latter  approach,  in  spite  of its shortcomings in defining the paths and
 measure rigorously, continues to be very useful. If the path integral is to
 be  treated  as an alternative procedure for quantization, one must  try to
 recover  at  least  all  the  exactly solvable problems of non-relativistic
 quantum  mechanics.  For  a  long  time,    apart from the quadratic action
 problems  the  only  non-gaussian  potential problem  which could be solved
 exactly  by  the  the  path  integral  method  was the $ 1/r^2 $  potential
 problem$^{11}.$  While  Duru  and  Kleinert$^{12}  $ succeeded in doing the
 exact  path integration for the Green function of the hydrogen atom problem
 using  the  Kustanheimo-Stiefel  transformation$^{13}  $, the manipulations
 used   in   the   original   paper  were  formal  and  lacked  mathematical
 justification. For the H- atom problem a complete treatment was given by Ho
 and   Inomata$^{14}.$   For   many  other  potential  problems  exact  path
 integartion  could  be  completed  in  the  next  few  years. This has been
 possible  thanks  to   the  idea of using local scaling of time in the path
 integral approach. Change of variable followed by local scaling of time  or
 addition  of new degrees of freedom have since been  used  by  many authors
 to   give   path  integral   solutions   for   many  potential problems  of
 quantum   mechanics.   The   Coulomb  problem  in   two dimensions$ ^{15} $
 and the radial  Coulomb  problem$^{16},$  the  Coulomb  field with magnetic
 charges$^{17}   $  and  the   Coulomb  potential  with  the  Bohm  Aharonov
 potential$^{18}  $  and  the non-relativistic dyonium problem $ ^{19} $ are
 some  of  the   Coloumb   field  related  problems which have  been  solved
 exactly   in   the   path  integral formalism. Important potential problems
 such as the Morse oscillator $^{20}, $ the Rosen-Morse potential $ ^{21}, $
 the   Poschl-Teller  oscillator $ ^{21},$ the Hartmann potential $ ^{22}, $
 the   Hulthen potential$^{24}, $ the $ \delta - $  potential$ ^{25}, $  the
 square  well  $ ^{26} $  and many  other  potential problems $ ^{27} $ have
 been solved exactly by the path integral method.

 Motivated  by  the importance of local scaling of time in the path integral
 formalism,  in  this  paper  we  introduce  a  new kind of hamiltonian path
 intergral  which  is termed as hamiltonian path integral with local scaling
 of  time.  This new path integral is defined in terms of  canonical form of
 the  path  integral   introduced  by one of us earlier $ ^{28} $. A special
 case  of  the  new path integral to be defined here has been found suitable
 for  hamiltonian   path   integral  quantization  scheme  in  arbitrary co-
 ordinates  $  ^{29}  $.  These  papers  will, hereafter, be  referred to as
 papers  I and II respectively. {\it The quantization scheme of paper II has
 the  merit  that, irrespective of the co-ordinates chosen, it is formulated
 in  terms  of  classical   Hamiltonian  only.} This may be contrasted  with
 existing Hamiltonian path integral quantization methods in which it becomes
 necessary  to  add   the  $  O(\hbar  ^2)   $ effective  potential  to  the
 classical Hamiltonian in non-cartesian co-ordinates$^{30}.$

 The  new  path integral intorduvced in this paper can be considered to be a
 generalization  of  the path integral defined in the paper I. In this paper
 we  study  in  detail the two hamiltonian path integrals of paper I and its
 generalization  defined  here  and  obtain  several important results about
 them.  Our  discussion  of  local  scaling  of  time  differs  from   other
 treatments   existing  in  the  literature  $  ^{31}.  $  In  the  existing
 treatments,  quantum  mechanics  is  assumed  and  the path  integral  with
 local   scaling   of   time   is  derived. However, we  directly define the
 canonical  path  integral  with  local   scaling  of time. We then obtain a
 connection   with   quantum   mechanics   by  deriving the time development
 equation  implied  by  the  assumed skeletonised form of the path  integral
 with   scaling.   For   path  integrals, without local scaling of time, the
 method  of   obtaining  the  Schrodinger equation is  well-known.  However,
 for   the  path integrals with local rescaling of time our approach is  the
 first of its kind for deriving the time development equation.

The  plan  of  this  paper  is  as  follows.  As  a  preparation to study of
hamiltonian  path  integration  with non-trivial scaling, in Sec. 2 we first
recall the definition of hamiltonian path integral of paper I and prove some
of  its  properties  needed later. The Schrodinger equation and the ordering
implied by the hamiltonian path integral are derived. The central results of
this  paper  on  properties of the hamiltonian path integral with nontrivial
scaling  are  derived  in  Sec.3.  As  applications  of  the results in this
section,  we  recover  the   quantization  scheme   of  paper  II.  The  two
dimensional  Coulomb  problem  in  parabolic  coordinates is discussed as an
example of quantization in arbitrary coordinates. This example is relatively
simple,  setting  up   path  integral quantization in arbitrary co-ordinates
immediately  leads  to  exact answer for the propagator. A detailed solution
for many other potential problems can be given using results derived in this
paper and will be taken up in a separate publication.

\vspace{5 mm}
\noindent
           {\bf 2. The hamiltonian path integral without scaling }
\vspace{5 mm}

{\it  2.1  Introduction  : }  In this section shall define hamiltonian  path
integral  without  scaling  and  discuss  its properties as a preparation to
study  of  hamiltonian  path  integartion with local scaling of time in Sec.
3.{\it  We  shall  assume  throughout  this  paper  that the Hamiltonian  is
independent of time and that it is quadratic  in  momenta.} To  simplify our
discussion  we  shall  not  keep  terms  linear in momenta, for these do not
require any technique other than that developed in our paper.

In the time slicing approach the basic building block of the path integral is
the  approximate expression of the propagator  for short times, which will be
called  the  short  time propagator (STP). {\it If the path integral is to be
used  as  an  independent  scheme  of  quantization,  the act of quantization
consists  precisely   in  choosing  the STP and using it to do summation over
histories.}   Let us generally denote the STP as $ (qt \vert q_0 t_0 ). $  As
it   stands   it  is merely a generalized function of the indicated variables
and,   for  short   times,  approximates  the  matrix  element  of  a unitary
transformation. For the Lagrangians of type
               \begin{equation}
                      L =\frac{1}{2} g_{ij} (q) \frac{dq^i}{dt} \frac{dq^j
                      }{dt} - V(q)
               \label{(4)}    \end{equation}
\noindent
an important form of STP is the Van Vleck- Pauli- Morette form  of short time
approximation $^{32,33,34} $ given by
             \begin{equation}
             (qt \vert q_0 t_0 ) = (2\pi i\hbar )^{-n/2} \bigl(g(q)g(q_0 )
             \bigr)^{1/4}
             \sqrt{ D}  \exp \bigl[ i S(qt,q_0 t_0 )/\hbar \bigr]
                                                           \label{(5)}
                                                      \end{equation}

          \noindent
Here  $  n  $  is  the number of degrees of freedom, $ S $ is  the classical
action  for  the  trajectory  passing through  the  points  $ (q,t) $  and $
(q_0  ,t_0  ),  g  =  \det(g_{ij} ) $ and $ D $ is the Van Vleck determinant
given by
             \begin{equation}
               D = \det \left( - \frac{\partial ^2 S }{\partial q^i
               \partial q^j_0 } \right)
                                    \label{(6)}
\end{equation}

          \noindent
This Van Vleck-  Pauli-  Morette  (VPM)  form  has  the following remarkable
properties  $^  {3,35}  $  which  will  be  generalized  when setting up the
hamiltonian path integration.
             \begin{enumerate}
\item{ In those cases  where  the  quantization can be done  using canonical
methods, the  semi-classical  limit  of  the  exact propagatoris seen  to be
just  the  VPM  formula,  even  for finite $ t-t_0 $. In some special  cases
the  semi-classical limit is already the exact propagator $ ^{35}. $ }

\item{   If    $  q_2  $  integral   is  computed  in  the  stationary  phase
approximation the VPM form of STP obeys the semi-group property  }

             \begin{equation}
             \int(q_3 t_3\vert q_2 t_2) \rho(q_2) d^n q_2 (q_2t_2\vert q_1
             t_1) =
               (q_3 t_3\vert q_1 t_1)           \label{(18)}
             \end{equation}

\item{  For $  t-t_0  \rightarrow 0 $ it obeys the boundary condition
              \begin{equation}
                     \lim_{t \rightarrow t_0} (qt \vert q_0 t_0 )  =
                     g(q)^{-1/2} \delta (q-q_0 )
          \label{(9.4)}   \end{equation}}

\item{  Under  arbitrary  co-ordinate   transformations  the  expression for
STP   remains  unchanged  because  both   the   classical trajectory and the
action  along  it   are   invariant   concepts.  Moreover  the normalization
(\ref{(9.4)}) is invariant  under  arbitrary co-ordinate transformations. }

\item{ The  action $ S $ is  the  generating  function   of   canonical
transformation connecting the  co-ordinates  and  momenta  at times $ t $ to
those at  times  $ t_0 $ .  It  satisfies  the  Hamilton Jacobi equation
              \begin{equation}
                \frac{\partial S}{\partial t}  +  H \bigl(q, \frac{\partial
                S}{\partial
                q}  \bigr)=0 \label{(8)}
          \end{equation}} \end{enumerate}

          \noindent
The  Lagrangian  form  of path integration in arbitrary  co-ordinates can be
obtained  by  using the VPM form of STP to do summation over histories. This
gives  the  correct  quantization  rules in  arbitrary co-ordinates when the
classical Lagrangian is used  $ ^{36} $.

{\it 2.2 The Hamiltonian Path Integral :} The  VPM  form  of  the short time
approximation  to   the propagator  shows  the  crucial significance  of
the classical trajectory and the classical action. Using this anology  we
shall now define the STP for the hamiltonian path integral  as a two step
process.

Let   $ \gamma _1 $ and $ \gamma _2 $  be two  classical  trajectories  with
boundary conditions as indicated below.

            \begin{equation} \gamma_1  : \tau   \rightarrow \bigl( \tilde
            {q}(\tau ),\tilde
            {p}(\tau )\bigr)  ,\  \ {\rm for} \ \   t_0 \leq \tau  \leq t_1
            \label{(9.6)}
            \end{equation}
            \begin{equation} \tilde {q}(t_0) = q_0 ,\ \ \  \tilde {p}(t_1)
            = p_1
            \label{(2.3)} \end {equation}
            \begin{equation}\gamma_2   : \tau  \rightarrow  \bigl( \tilde
            {\tilde {q}}(\tau
            ),\tilde {\tilde {p}}(\tau )\bigr) ,\ \  {\rm for} \ \  t_1\leq
            \tau  \leq t \label{(9.8)}
            \end{equation}
            \begin{equation} \tilde {\tilde {p}}(t_1) = p_1 ,\ \ \   \tilde
            {\tilde {q}}(t)
            = q         \label{(2.4)} \end{equation}
            \begin{equation} (qt\vert p_1t_1)   = (2\pi \hbar )^{ - n/2}
             \sqrt{D_{++}} \exp [ i S_{++}(qt,p_1t_1)/
              \hbar] \label{(2.5)} \end{equation}
            \begin{equation}
            (p_1t_1\vert q_0t_0) = (2\pi\hbar )^{-n/2} \sqrt{  D_{--}}\exp
            [
            i S_{--}(p_1t_1,q_0t_0)/\hbar ]               \label{(2.6)}
            \end{equation}

           \noindent
            where
            \begin{equation} D_{++}= \det \left(\frac{ \partial^2 S_{++}}{
            \partial q^i \partial p_{1j}}\right)
                                      \label{(2.7)}        \end{equation}
            \begin{equation} D_{--} = \det \left( \frac{\partial ^2 S_{--
            }}{ \partial q^i_j
            \partial p_{1j}} \right) \label{(2.75)}      \end{equation}

            \noindent
and $ S_{++} , S_{--}  $ are Legendre transforms of the classical action
along the two trajectories,

            \begin{equation} S_{++}(qt,p_1t_1)  =  p_{1i} \tilde {\tilde
            {q}}_1^{~i}  +
            \int_{\gamma_2} ( \tilde {\tilde {p}}_{~i} d \tilde {\tilde
            {q}}^i(\tau )- H(\tilde
            {\tilde {q}}(\tau ),\tilde {\tilde {p}}(\tau))) d\tau
            \label{(2.9)}
            \end{equation}
            \begin{equation} S_{--}(q_0t_0,p_1t_1) = -p_{1i} \tilde
            {q}_1^{~i} +
            \int_{\gamma_1}(\tilde {p}_id\tilde {q}^i- H(\tilde
            {q}(\tau),\tilde
            {p}(\tau)))\  d\tau  \label{(2.10)} \end{equation}

           \noindent
           where
            \begin{equation} \tilde {q}_1 \equiv   \tilde {q}(t_1), \ \ \ \
            \ \tilde
            {\tilde { q}}_1 \equiv   \tilde {\tilde { q}}(t_1)
            \label{(2.11)}\end{equation}

In (\ref{(2.5)}) to (\ref{(2.10)}) it is understood that the independent
variables  are those explicitly shown on the left hand side. Also the right
hand sides are supposed to  have  been  expressed  in  terms  of  these
variables using canonical transformations, or equivalently,  using the
equations for the  classical  trajectories  (\ref{(9.6)})  and
(\ref{(9.8)})  as functions of boundary conditions.

The canonical STP is now defined as
          \begin{equation} (qt\Vert q_0t_0) = \frac{1} {\sqrt{\rho (q)\rho
          (q_0)}} \int d^n p_1 ~ (qt\vert p_1t_1) ~ (p_1t_1\vert q_0 t_0)
            \label{(2.12)}
            \end{equation}
            \noindent
where  we  have  introduced  the factor $  \sqrt{\rho(q)\rho(q_0)}  $ in the
definition  of  STP  so  that  it  satisfies  the semi-group property in the
stationary phase approximation with respect to the measure $\rho (q) d^n q.$
             \begin{equation}
             \int(q_3 t_3\Vert q_2 t_2) \rho(q_2) d^n q_2 (q_2t_2\Vert q_1
             t_1) \approx
               (q_3 t_3\Vert q_1 t_1)           \label{(9.18)}
             \end{equation}

             \noindent
Several remarks are now in order concerning the above definition.
             \begin{enumerate}

               {\item The functions $ S_{++} $ and $ S_{--} $ obey  the
               corresponding
                Hamilton Jacobi  equations.
               \begin{equation}
                       \frac{\partial S_{++}}{\partial t}    +  H \left( q,
                       \frac{\partial S_{++}}{\partial q} \right)  = 0
                                    \label{(19)}
          \end{equation}
          \begin{equation}
                        \frac{\partial S_{--}}{\partial t_0} +H
                        \left( \frac{\partial S_{--}}{\partial p},p \right)
                        = 0
                         \label{(20)}
          \end{equation}}

{\item  The functions $ S_{++} $ and $ S_{--} $ always exist  for  any given
$  H(q,p)  $  because   the  variables  $ (q,p_1), $ and also $(q_0, p_1), $
always  form  independent  sets  of  variables,  at  least  for small enough
intervals  $  t-t_1  $  and   $t_1-t_0 $. The same cannot be  said  for  the
pairs  $  (p,p_1)  ,(q_2,q),(q,q_1)  \  {\rm  or  }\  (p_1,p_n) $.  If,  for
example,  $  H(q,p)  $   is  a  function of $ p' $  s alone, the momenta are
constants  of  motion  and,  therefore,  the values of momenta at  different
times cannot be independent.}

{\item The canonical STP satisfies the boundary condition
          \begin{equation}
                        \lim_{ t \rightarrow t_0 } (qt\Vert q_0 t_0 ) =
                        (\rho(q))^{-1/2} \delta^{(n)}(q-q_0 )  \label{(9.21)}
                        \end{equation}
Such  a  STP,  after  doing  a summation over histories, will give rise to a
unitary transformation  on  a  Hilbert space  with measure $\rho(q). $ }

{\item  The  expression (\ref{(2.12)}) becomes identical with the  VPM  form
if  the   $  p_1-  $   integrals  are  evaluated  in  the  stationary  phase
approximation. }

{\item  The  expressions  $ (qt\Vert p_1t_1) $ and $ (p_1t_1\Vert q_0t_0 ) $
correspond  to  short  time   approximations,  respectively,  of the quantum
mechanical amplitudes.
               \begin{equation}
                       <q \vert  \exp [-iH_{op}(t-t_1)/\hbar ]  \vert
                       p_1> \end{equation}
          and
          \begin{equation}
                   <p_1 \vert  \exp [-i H_{op}(t_1-t_0 )/\hbar ]
                   \vert q_0 >
                                         \end{equation}   }
             \end{enumerate}

We  shall  now  discuss  the  properties of the hamiltonian path integral of
paper  I obtained by summing over histories with (\ref{(2.12)}) inserted for
STP.  We  shall derive the Schrodinger equation satisfied by the hamiltonian
path  integral  and  the operator ordering rule which relates the operator $
H_{op}  $  appearing in the above  expressions  to  the function  $ H(q,p) $
used to set up the STP.

The STP, as defined by (\ref{(9.6)})-(\ref{(2.12)}) above, will hereafter be
called  the  {\it  canonical STP }  for the Hamiltonian function $ H(q,p). $
It gives rise to a path integral in the limit $  N \rightarrow \infty $ of
                  \begin{equation}
                K ^{(N)}[H,\rho](qt;q_0 t_0 ) = \int \prod^{N-1} _{k=1}
                \rho (q_k )dq_k
                 \prod ^{N-1}_{j=0} (q_{J+1} \epsilon \Vert q_j 0)
               \label{(9.23)}
               \end{equation}
where   $ \epsilon = (t-t_0 )/N $ and $ q_N \equiv q.$  The path integral so
obtained   will be referred to as the {\it first hamiltonianl path integral,
or  HPI1,}  for  the   Hamiltonian  function  $ H(q,p) $ and  measure $ \rho
(q).$  It  will  be   denoted   by  $ K[H,\rho ](qt;q_0 t_0 ). $  Thus
             \begin{equation}
                   K [H,\rho ] (qt;q_0 t_0 ) =   \lim _{N \rightarrow
                   \infty }
                   K^{(N)}[H,\rho ](qt;q_0 t_0 )       \label{(9.24)}
                                         \end{equation}
{\it 2.3 Elementary properties of the hamiltonian path integral : }

(a)  It  is  useful  to  note here that, if a constant $ E $ is added to the
hamiltonian, the HPI1 changes by a phase factor

             \begin{equation}
               K[H-E,\rho ](qt;q_0 t_0 ) = \exp[iE(t-t_0 )/\hbar ] K[H,\rho
               ](qt;q_0 t_0)
             \end{equation}

(b) The dependence of the HPI1  $ K[H,\rho ] (qt;q_0 t_0 ) $ on the measure
$ \rho $  is very simple. In fact we have the following relations
             \begin{eqnarray}
             \sqrt{\tilde{\rho} (q) \tilde{\rho }( q_0 )} K[H,\tilde{\rho}]
             (qt;q_0 t_0 )
              &  = & \sqrt{ \rho (q)\rho (q_0 )} K[H,\rho ] (qt;q_0 t_0 )
              \nonumber \\
                                         & = &{\rm  independent \ of \ }
                                         \rho
                                                 \label{(24)}
               \end{eqnarray}

          \noindent
To  prove  the  above  statement,  observe  that  in  the  expression  for $
K^{(N)}[H,\rho  ]  $  the  factors  $  \rho (q_1)...\rho (q_{N-1})  $ coming
fromthe   STP  cancel with  those  from  the  integration  measure   leaving
the  factor  $ \sqrt{ \rho (q)\rho (q_0 )} $ coming from $ \rho $ on the two
end points only. Thus $ K^{(N)}[H,\rho ] $ and $  K^{(N)}[H,\tilde{\rho}]  $
constructed   for   different   measures  satisfy  a  relation  similar   to
(\ref{(24)})  and the proposition follows by taking the $ \lim N \rightarrow
\infty. $

(c)  In  the  cartesian coordinates $ K [H,\rho] $ gives the correct quantum
mechanical propagator.

{\it 2.4 Discretized form for the hamiltonian path integral :}
The  hamiltonian  path   integral $ K[H,\rho ] $  defined  by (\ref{(9.23)})
and   (\ref{(9.24)}),   represents   the  matrix   element  of   a   unitary
transformation.  Therefore  the  wave  function  propagated  by it satisfy a
certain Schrodinger equation. In order to derive the Schrodinger equation we
need an explicit expression for the STP $ (qt \Vert q_0 t_0 ).$

{\it Proposition 2.1 :} Let $ \epsilon _1 = t-t_1 $ and $ \epsilon _2 = t-
t_0 $  be small. Then
                  \begin{eqnarray}
                       S_{--}(p_1t_1 ,q_0 t_0 ) & = &- p_{1i} q^i_0  -
                       H(q_0 ,p_1)
                       \epsilon _1 + O(\epsilon ^2_1 )
                       \\
                       S_{++}(qt,p_1t_1) &  = &  p_{1i}q^i    - H(q ,p_1)
                       \epsilon_2
                       + O(\epsilon ^2_2 )     \label{(26)}
                       \end{eqnarray}
The proof is fairly straightforward computation using the  regular behavior
of the classical trajectories $ \gamma _1 $ and $ \gamma _2 $ and formulas
such as
             \begin{eqnarray}
             \int^{x_2}_{x_1}  f(x) dx &  = & \sum _{n=0}^{\infty}
             \frac{1}{(n+1)!}
             f^{(n)} (x_1)  (x_2-x_1)^{n+1}
              \label{(27a)}
     \\ \int^{x_2} _{x_1} f(x) dx & = & \sum
                         _{n=0}^{\infty}
                         \frac{(-1)^n}{(n+1)!} f^{(n)}(x )(x_2-x_1)^{n+1}
     \label{(27b)}
     \end{eqnarray}

Applying this result to ( \ref{(2.7)})-(\ref{(2.10)}) we immediately get

{\it Proposition 2.2:} For small  $ \epsilon _1  $ and $ \epsilon _2 $ we
have
               \begin{eqnarray}
               (q\epsilon\Vert q_0 0) =
                \bigl( \rho (q)\rho (q_0 ) \bigr)^{1/2} \int
               \frac{d^np}{(2 \pi \hbar)^n} \exp [ip_i(q^i  -q^i_0)/ \hbar
               ] \times
          \hspace{1.1 in} \cr
               \exp \bigl[ -i H(q,p)\epsilon/\hbar  -i H(q_0 ,p)
                                 \epsilon/ \hbar  \bigr] \left[ 1 -
                                 \frac{\epsilon_1}{2}
               \frac{\partial ^2 H(q,p)}{\partial q^i \partial p_i}
           + \frac{\epsilon_2}{2} \frac{\partial ^2 H(q_0,p)}{\partial q^i_0
           \partial p_i}
               + \cdots   \right]   \label{(51)}
                                \end{eqnarray}
                                 \noindent

This is the form which will be useful for later aplications. For hamiltonian
quadratic  in  momenta,  the  hamiltonian  STP  can be reduced to a standard
Lagrangian  form  of STP, such as the mid point form, by carrying out the p-
integrations  and  using the McLaughlin Schulman trick ${^36}.$ It is fairly
easy  and  straight  forward to work out and the details can be found in ref
37.  For  the present purpose we write it in a slightly different form valid
for  short  times  From  now  onwards  we  shall choose $ t_1 $ to be midway
between $ t_0  $ and $ t.$ Thus we shall take
          \begin{equation}
              t-t_1 = \epsilon_2 = \epsilon_1 = t-t_0  = \epsilon
              \label{(29)}
                                          \end{equation}

\vspace{.3 cm}

\noindent
The reason for  this  choice  is  that  it  gives  a  self- adjoint (symmetric)
Hamiltonian  operator,  though  for $ H $ quadratic  in momenta
it is not essential to make this choice. Neglecting $O(\epsilon^2) $ terms we
get
                \begin{eqnarray}
        \lefteqn{    (q\epsilon\Vert q_0 0) } \nonumber
                   \\ && = \bigl(\rho (q)\rho (q_0
                     )\bigr)^{1/2} \int
                     \frac{d^np}{(2 \pi \hbar)^n} \exp[ip_i(q^i  -q^i_0)/
                     \hbar ] \times \nonumber
             \hspace{1.2 in} \\  &&
            \left[ 1 - \frac{i\epsilon }{\hbar} H(q,p) -\frac{i
            \epsilon}{\hbar}H(q_0 ,p)
               - \frac{\epsilon}{2} \frac{\partial ^2 H(q,p)}{\partial q^i
               \partial p_i} +
              \frac{\epsilon}{2}\frac{\partial ^2 H(q_0,p)}{\partial q^i_0
              \partial p_i}
             + ...    \right]
                                              \label{(28)}
                                  \end{eqnarray}
             \noindent
{\it  2.5  Schrodinger equation for the hamiltonian path  integral.}  Let us
consider the Hilbert  space  of  wave  functions  $ \psi(q) $ defined on the
configuration space with measure $ \rho (q)d^n q . $ In  other words, $ \psi
\in   L  ^2(R^n,\rho (q) d^nq) $  if $\int  \vert \psi (q) \vert ^2 \rho (q)
d^n q \leq \infty . $ Let $ \psi $  be propagated by the canonical STP :

             \begin{equation}
           \psi (q,t) = \int K[H,\rho ](qt;q_0 t_0 ) \psi (q_0 ,t_0 ) \rho
           (q_0 )
            d^n q_0   \label{(30)}
                  \end{equation}

                  \noindent
The equation  satisfied  by $ \psi (q,t) $  is  given  by  the  following
proposition.

{\it Proposition 2.3 : } $ \psi (q,t) $ defined by (\ref{(30)}) satisfies
                          \begin{equation}
                i\hbar \frac{\partial \psi}{\partial t}(q,t) = H_{op}
                \psi (q,t)
                                 \label{(31)}
\end{equation}
where  the  operator  $  H_{op},$   defined on $ L^2(R^n,\rho (q)d^nq),$ is
obtained  by the following prescription:

             \begin{enumerate}
              {\item Define
                       \begin{equation}
                       H_1(q,p) = \left[ H(q,p) - \frac{i\hbar
                       }{2}\frac{\partial ^2 H}{\partial
                        q^i \partial p_i} \right]         \label{(32)}
                                              \end{equation}}

 {\item  Bring  all   $ q' $ s in $ H_1 $ to the left and all $ p'$ s to the
 right. Replace each $ p_i $ in $ H_1,$  so ordered, by $ -i\hbar \partial /
 \partial  q^i.$  This  makes  $  H_1  $  a  differential  operator,  say, $
 X(q,\partial/ \partial q ), $ acting on functions of $ q. $           }

 {\item Next bring all  $ q' $ s in $ H_1 $ to the right and all $ p' $ s to
 the left. Replace each $ p_i $ in $ H_1,$ so ordered, by $ -i\hbar
 \partial/ \partial q^i. $  This makes $H_1 $ a differential operator, say,
 $ Y(q,\partial / \partial q), $ acting on functions of $ q. $           }

 {\item Define  $ \widehat{H} $  by symmetrization,
           \begin{equation}
            \widehat{H} = X \Bigl(q,\frac{\partial}{\partial q}\Bigr)
                  + Y \Bigl(q,\frac{\partial}{\partial q}\Bigr)
                                 \label{(34)}
              \end{equation}  }

           {\item Finally, define $ H_{op} $ by
           \begin{equation}
            H_{op} = \rho^{-1/2} \widehat{H} \rho^{1/2}
            \end{equation} }
             \end{enumerate}
                 \noindent
The  proof  of  this prescription is based on the  formula ( \ref{(28)}) and
the fact that, with $  x^i= (q-q_0)^i $,

          \begin{eqnarray}
          \lefteqn{ \int \rho (q_0 ) d^n q_0  \int \frac{d^np }{(2\pi
              \hbar)^n}
                     \exp (i p_i x^i / \hbar)
                     \frac{ f(q) p_j g(q_0) }{\sqrt{ \rho (q_0) \rho (q) } }
                       \psi (q_0 )
               }    \nonumber \\
                &&  =  f(q) \bigl(\rho (q)\bigr)^{-1/2} \int
                d^n q_0 \bigl(\rho (q_0 )\bigr)^{1/2}
                \psi (q_0 ) \bigl[ -i\hbar  \frac {\partial}{\partial x^j}
                \bigr]
               \delta ^n(x) \hspace{2 cm}      \nonumber
             \\ &&
               =   \left. f(q) \bigl(\rho (q) \bigr)^{-1/2} (-i\hbar
              \frac{\partial}{\partial q^j_0} )
              \bigl(  (\rho (q_0 ))^{1/2} g(q_0) \psi (q_0 ) \bigr)
              \right| _{q_0=q } \nonumber
             \\ &&
          =  (f(q) \hat{p}_j g(q) )\psi (q)  \hspace{3.5 in}  \label{(38)}
              \end{eqnarray}
          \noindent
The  expression   $  f(q)  p_j  g(q_0) $  inside the above integral becomes
equal to the or\-der\-ed op\-er\-a\-tor $ f(q) \hat{p_j} g(q). $  The above
rules  are  now  a  direct  generalizations  of these formulas to arbitrary
functions  $  H(q,p) $, at least, to  the  functions  which are polynomials
in  $  p.  $  For   real $ H(q,p),$  the operator $ H_{op} $ defined by the
proposition 2.3 is also hermitian on $ L(R^n,\rho(q)d^nq ) $

An  important  case  of  interest  is  the  Schrodinger  equation  when the
hamiltonian function is given by
     \begin{equation}
            H_0 =  \frac{1}{2m} g^{ij} p_i p_j
                                             \label{(40)} \end{equation}
 where $ g^{ij} $ is the contravariant metric tensor ( matrix inverse of
 $ g_{ij}  ).$  In this case we have
 \begin{eqnarray}
     \widehat{H_0}  & = & \frac{1}{2m} \frac{\partial}{\partial q^i}
                g^{ij}\  \frac{\partial}{\partial q^j}
                                                         \label{(41)}  \\
     (H_0)_{op} & = & \frac{1}{\sqrt{\rho}}\widehat{H_0}\sqrt{\rho} \\
              & = & - \frac{\hbar ^2}{2m} \frac{1}{\sqrt{ \rho}}
                 \frac{\partial}{\partial q^i} g^{ij} \left(
                   \frac{\partial}{\partial q^j} \sqrt{\rho} \right)
                         \label{(42)} \\
             & = &  - \frac{\hbar^2}{2m} \Delta _{\rho}    +  U (\sqrt{g})
               \label{(9.43)}
             \end{eqnarray}
Here  $  \Delta_\rho   $  is  self-adjoint  Laplace-Beltrami  operator on $
L^2(R^n  \rho  (q)d^n q) $  related to the normal Laplace-Beltrami operator
on $ L^2(R^n\sqrt{g} d^n q) $ as
             \begin{eqnarray}
                \Delta_{\rho} & = & \rho^{-1/2} g^{1/4} \Delta g^{-1/4}
                \rho^{1/2}
                                           \label{(46)}
             \\
                \Delta &  = & g^{-1/2}\frac{\partial}{\partial q^i} (
                g^{1/2}g^{ij}
                \frac{\partial}{\partial q^j} )
\end{eqnarray}
\noindent
with
             \begin{eqnarray}
               U(h) & = &-\frac{\hbar^2}{8m} \bigl[ g^{ij}(\ln h),_i(\ln
               h),_j + 2
               \bigl( g^{ij} (\ln h),_i \bigr),_j \bigr]
               \label{(44)}
               \\
                  & = & - \frac{\hbar^2}{2m} [ g^{ij} (\sqrt{h},_{ij} /
                  \sqrt{h}
                  + g^{ij}_i(\sqrt{h}),_j/ \sqrt{h} ]
\label{(45)}
\end{eqnarray}

The result (\ref{(9.43)}) shows that HPI1 $K[H,\sqrt{g}]$  for the classical
hamiltonian  $  H_0,$   set  up  as  in  paper  I, gives correct Schrodinger
equation  only in the cartesian co-ordinates and in general $ U(\sqrt{g}), $
which  is  of  the order $ \hbar ^2, $ must be subtracted from the classical
hamiltonian  to  get  correct  quantization schme.In this respect HPI1 is no
better than any other existing hamiltonian path integral scheme. However, as
we  shall  see in the next section, HPI1 can be used to define a second kind
of  path integral which involves suitable scaling of time and which with the
classical  hamiltonian  as  input  gives  rise  to  the  correct Schrodinger
equation in every co-ordinate system.


{\bf 3. Hamiltonian path integral with scaling}
\vspace{5 mm}

Local  scaling  of  time  has  proved to be a useful technique in completing
exact path integration for several potential problems. Just as in mathemaics
good  results  are  frequently turned into definitions, we use an expression
appearing  in  the scaling relation to define a hamiltonian path integral of
second  kind,  one with local scaling of  time. A special case of this  path
integral  already used in paper II which, with the classical hamiltonian and
a  suitable   scaling  function,  was  found  to  give  rise  to the correct
Schrodinger equaion in every set of co-ordinate system.

We  introduce  a hamiltonian path integral ${\cal K}[H,\rho,\alpha] $ with
scaling by means of the equation
             \begin{eqnarray}
             \lefteqn{  {\cal K}[H,\rho,\alpha] (qt,q_0 0)    } \nonumber
            \\  &&  \equiv \sqrt{\alpha (q)\alpha (q_0)}
             \int \frac{dE}{2 \pi \hbar} \exp(-iEt/\hbar) \int d\sigma
             K[\alpha (H-E),\rho]
             (q\sigma ,q_00) \hspace{.2 in}
                                                            \label{(72)}
                                                           \end{eqnarray}

\noindent
This  path  integral will be called the second hamiltonian path integral, or
HPI2.  It depends on the hamiltonian $ H(q,p), $  integration measure $ \rho
$  and scaling function  $ \alpha (q). $ In the right hand side of the above
equation $ K[\alpha (H-E),\rho] $ is the HPI1 obtained, after summation over
histories,  from  the  canonical  STP corresponding to the function $ \alpha
(H(q,p)-E).  $  The  above  definition  is  such  that  for $\alpha(q) = $ a
constant,   the  HPI2  $  {\cal  K}[H,\rho,\alpha]  $  coincides  with  HPI1
$K[H,\rho].$ We have found it useful to not to fix form of $ \alpha(q) $ and
$ \rho(q) $ at this stage just as the form of hamiltonian function $H(q,p) $
is  not  fixed  while defining the path integral. As we shall see below, the
path  integral  HPI2  generalizes  HPI1 of paper I and HPI2  reduces to HPI1
when $ \alpha(q) $ is indepndent of q.

The central results of this paper on local scaling of time in canonical path
integral are given in the propositions 3.1 to 3.3 given below.

{\it  Proposition  3.1:}  The  Schrodinger equation  satisfied  by  the wave
function  propagated  by  the  HPI2,  ${\cal  K} [H,\rho,\alpha] $ for $ H_0
=(1/2m)g^{ij} p_i p_j $ is

     \begin{equation}
      i\hbar \frac{\partial \psi}{\partial t}
      ={\rho}^{-1/2}(\alpha ^{-1/2}\widehat{\alpha H_0 }\alpha^{-1/2})
       {\rho}^{1/2} \  \psi  \label{(77)}
                          \end{equation}
\noindent
and $ {\cal K} [H,\rho,\alpha]$ has the normalization  $ {\rho}^{-1} \delta
(q-q_0).$In the above equation $\widehat{\alpha H_0 }$ is the operator
obtained from the function $\alpha(q) H_0(q,p)$ by applying rules 1 to 4
given in proposition 2.3.

We  shall  prove  the  proposition  for  $\rho  = 1  $ ;  there will be  no
$\sqrt{\rho(q_k)   \rho   (q_{k+1})   }   $    factors   which  occur   for
normalization   corresponding   to $ \int \rho d^n q $ in (\ref{(51)}). The
proof  for  an  arbitrary  $  \rho  $  does  not   involve   any additional
difficulty.

We   now  construct the canonical STP from (\ref{(51)}) and set up its $ N-
$  fold  integration;  there are $2N-1$ intermediate points in the interval
$[0,\sigma]$      and     and     corresponding     variables     are     $
p_0,q_1,p_1,...,q_{N-1},p_{N-1}.$ We call  $q_N.$ the end point $q.$
             \begin{eqnarray}
         \lefteqn{
             K^{(N)}[\alpha(H-E),1](q\sigma ,q_00) \hspace{3.8 in}
             } \nonumber \\
             && = \int dp_0 \prod^{N-1}_{r=1}\left(\int
             \frac{dq_r dp_r}{(2 \pi \hbar)^n} \right)
             \exp\left[ \frac{i}{\hbar} \sum^{N-1}_{j=0}
             S_E(q_{j+1},p_j,q_k;\sigma )\right]     \times  \hspace{0.9 in}
             \nonumber
             \\ && \hspace{1.0 in}
                 \prod^{N-1}_{k=0}  \left[  1
                  -\frac{\sigma}{4N}p_{ki}
                  \frac{\partial ( \alpha _{k+1} g^{ij}_{k+1})}{\partial
                  q^j_k}
                  +\frac{\sigma}{4N}p_{ki}
                   \frac{\partial ( \alpha _k g^{ij}_k)}{\partial q^j_k}
                   \right] \                     \label{(75)}
\vspace{5 mm}                          \end{eqnarray}
where

             \begin{eqnarray}
             S_E (q_{k+1},p_k,q_k;\sigma )  \hspace{4.0 in}  \cr
               =  p_{ki}(q^i_{k+1} - q^i_k) - \frac{\sigma}{2N}
               \big(\alpha_{k+1}
             H_0(q_{k+1},p_k) + \alpha _kH_0(q_k,p_k) -(\alpha
             _{k+1}+\alpha _k)E \bigr)
                               \label{(76)} \end{eqnarray}
and
             \begin{equation}
                \alpha _k  =  \alpha (q_k), \ \ \    g^{ij}_k = g^{ij}(q_k)
            \ \ \, q_N = q  \hspace{2 cm }{\rm etc.}
          \label{(553)}                          \end{equation}

             \noindent
and we hope that the pseudo-time  lattice  index $  k $ will  not  be
confused with the indices $ i $ and $ j $ referring  to  the  different
co-ordinates.  In  any  case  all  summations  and  products  are displayed
explicitly.

Substituting the expression (\ref{(75)}) for $ K^{(N)} in ,$
      \begin{eqnarray*}
       {\cal K}[H,1,\alpha ](qt,q_00) \hspace{3.7 in} \nonumber
       \end{eqnarray*}
       \begin{equation}
              \equiv\lim_{N \rightarrow \infty} \sqrt{\alpha (q)\alpha
         (q_0)}\int_{- \infty}^{\infty} \frac{dE}{2 \pi \hbar} \exp(-iEt/\hbar)
               \int_0^{\infty} d\sigma  K^{(N)}[\alpha (H-E),1](q\sigma
               ,q_0 0)     \nonumber    \\          \label{(555)}
             \end{equation}
\noindent
we first evaluate the $ E $ integral which gives us the delta function
            \begin{equation}
                \delta  \Bigl( \frac{\sigma}{2N}( \alpha_0   + 2
                 \sum ^{N-1}_{k=1}\alpha _k +\alpha ) - t \Bigr)
                 \label{(79)}
                   \end{equation}
             \noindent
where   $  \alpha_N  =  \alpha(q)  =  \alpha,  \alpha_0=\alpha(q_0) .$ This
enables  us  to  do  the $ \sigma $  integral. Since all $ \alpha ' $ s are
positive,  the  $  \delta  -  $ function definitely contributes for  some $
\sigma  $ in  the  range $ (0,\infty ).$ In fact it contributes a factor
           \begin{equation}
                 2N \Bigl( \alpha _0 + \alpha _N +2 \sum^{N-1}_{k=1}\alpha
                 _k
                 \alpha _k \Bigr)^{-1} \equiv  2N / F(q_0,q_1,...,q_N)
                 \label{(80)}
                 \end{equation}
at
             \begin{equation}
                                 \sigma   = 2Nt / F(q_0,q_1,...,q_N)
                                 \label{(81)}
                                 \end{equation}
\noindent
 For later use we note that when all the  arguments  $ q_0,...,q_N $ are equal
  $ F/2N $ becomes equal to  $ \alpha (q), $ i.e.,
             \begin{equation}
                       F(q,q,...,q) / 2N = \alpha (q)
                       \label{(82)}
                    \end{equation}

 \noindent
Performing the $ \sigma $ integral with the help  of  the $  \delta -  $
function  in (\ref{(79)})  as explained above gives
\pagebreak
             \begin{eqnarray}
            \lefteqn{{\cal K}[H,1, \alpha ](qt,q_00) \hspace{4.7 in} }
               \nonumber
            \\ &&  = \lim_{N \rightarrow \infty} \sqrt{\alpha (q)\alpha (q_0)}
              \int dp_0
             \prod^{N-1}_{r=1} \left(  \int \frac{dq_r dp_r}{(2 \pi
             \hbar)^n}\right)
             \bigl(\frac{2N}{F} \bigr)  \times  \hspace{1.5 in}  \nonumber
            \end{eqnarray}
            \begin{equation}
             \exp\Bigl[ \frac{2i}{\hbar}  \sum^{N-1}_{j=0}
             S(q_{j+1},p_j,q_j;\frac{2Nt}{F}\Bigr]
              \prod^{N-1}_{k=0}  \Bigl[1-  \frac{t}{2F} p_{ki}
                      \frac{\partial (\alpha_{k+1} g^{ij}_{k+1})}{\partial
                      q^j_{k+1}}
                      +\frac{t}{2F} p_{ki}
                      \frac{\partial (\alpha_k g^{ij}_k)}{\partial q^j_k }
                      \Bigr]
             \nonumber \\ \label{(83)}
                      \end{equation}
Our  aim is to find the Schrodinger equation satisfied by  the wave function
propagated by  $ {\cal K}[H,1,\alpha ]. $  For this purpose it is sufficient
to  retain  the  terms  of order $  t $ and neglect  all  other higher order
terms.  We  must  emphasize  here  that  we  cannot take $\sigma$ small  and
replace  HPI1  in  (\ref{(72)})  by  a {\it single } STP. For small $ t $ We
write the factors  under the  integral sign in (\ref{(83)} as
           \begin{eqnarray}
          \lefteqn{
          \frac{2N}{F} \exp \big[ \frac{i}{\hbar} \sum ^{N-1}_{j=1}
          p_j(q_{j+1}-q_j)
           \bigr] \times  \hspace{1.5 in}
           } \nonumber
           \\  &&
            \left\{  1 - \frac{it}{\hbar F} \sum ^{N-1}_{m=0}
           \bigl( \alpha _{m+1}H_0(q_{m+1},p_m) + \alpha _mH_0(q_m,p_m)
           \bigr)\right\}
          \times    \nonumber
          \\&&  \hspace{2. cm} \left\{
             1 - \frac{t}{2F} \sum^{N-1}_{m=0} \left[
            \frac{\partial}{\partial q^i _{m+1}}
          \Bigl(\alpha _{m+1}g^{ij}_{m+1}\Bigr)\ p_{mj}-
          \frac{\partial}{\partial q^i_m}
           \Bigl(\alpha _m g_m^{ij} \Bigr) \  p_{mj} \right]   \right\}
           \nonumber \\
            \label{(84)}
                     \end{eqnarray}

\noindent
For the terms independent  of $ t,$  i.e., in  the  leading term,
$1,$ coming from the product of the expressions inside the curly brackets,
there is no $ p- $  dependence at all and the $ p $ integrations give rise to a
product  of $  \delta - $
functions in differences $ q_{k+1}-q_k $ which  can  be  used  to  perform
all $  q_k, k=1,2,...,q_{N-1}, $ integrals. Noting (\ref{(82)}) we get
          \begin{equation}
           \sqrt{\alpha (q)\alpha (q_0)} \Bigl[\frac{2N}{F}\Bigr] \delta
           ^n(q-q_0)
           = \delta ^n(q-q_0)               \label{(85)}
                                             \end{equation}

             \noindent
Next we consider the terms of order $ t  $ in  (\ref{(84)}). These  terms
are typically of the form
             \begin{eqnarray}
             \sqrt{\alpha (q) \alpha (q_0)} \int dp_0 \left( \prod ^{N-
             1}_{r=1}
              \int \frac {dq_r dp_r} {(2\pi \hbar)^n} \right)
          \Bigl(-\frac {2iN} {\hbar F^2} \Bigr) \times  \hspace{4. cm}
             \cr \hspace{1 in}
                \exp \Bigl[\frac{i}{\hbar}\sum_{j=1}^{N-1}
                  p_k(q_{j+1}-q_j)
          \Bigr] J(q_{m+1},p_m,q_m)  \hspace{2. cm}
          \label{(86)}
                  \end{eqnarray}

\noindent
with $  J $ as an abbreviation for the $ m^{th} $ term obtained after
multiplying out the curly brackets in (\ref{(84)}). In the $  m^{\rm th} $
term in (\ref{(84)}) all but the $ p_m $ integral  can  be done trivially to
yield
             \begin{eqnarray}
          \lefteqn{   \sqrt{\alpha (q)\alpha (q_0)} \bigl(-
          \frac{it}{\hbar}  \bigr)
          (2N) \Bigl((2m+1)\alpha_0
              + (2N-2m-1)\alpha \Bigr)^{-2} \times \
              }    \nonumber  \\
           &&  \int \frac{dp}{(2 \pi \hbar)^n} \exp \bigl(ip_m(q-q_0)/
           \hbar \bigr) J(q,p_m,q_0)  \hspace{1.5 in}  \nonumber
             \\
             &&      =(-i2tN/\hbar)\int^{\infty}_0  \beta d\beta
             \sqrt{\alpha (q) }
             \exp \bigl(-\beta (2N-2m-1)\alpha (q)\bigr) \ \ \times
               \nonumber \\
            &&   \left[ \int \frac{dp}{(2 \pi \hbar)^n} \exp \bigl(ip(q-
            q_0)/\hbar\right)
               J(q,p,q_0)\Bigr] \sqrt{\alpha (q_0) }  \exp \bigl(-\beta
               (2m+1)\alpha_0
               (q)\bigr)  \nonumber
               \\ \label{(87)}
               \end{eqnarray}
             \noindent
In the last step  above  we  have  dropped  the  subscript $ m $ from
the integration variable $p_m$  and  have  replaced  the  factor
$ (...)^{-2} $ by  an integral over a $ \beta $ using the identity

             \begin{equation}
               \frac{1}{x^2}  = \int_{0}^{\infty} \beta  d\beta  \exp(-
               \beta x)
                                    \label{(88)}
\end{equation}

            \noindent
We have also ordered the $ q  $ and $ q_0 $ dependence to  the  left  and to
the  right.  Now  noticing  that  $  J(q,p,q_0)   $  is exactly the  order $
\epsilon  (  = t \ {\rm here}) $ term in (\ref{(38)}) for $ \widehat{(\alpha
H_0)}   $  and   proceeding   in   manner  identical   to   the   proof   of
proposition 2.3, we realize that in the action  of ${\cal K}(qt,q_0 0)  $ on
$ \psi (q_0) $
this term contributes
          \begin{eqnarray}
             - \Bigl(\frac{2iNt}{\hbar}\Bigr) \int^{\infty}_0  \beta
             d\beta
             \sqrt{\alpha(q)}
              \exp \bigl[-\beta (2N-2m-1)\alpha (q) \bigr]  \times
              \widehat{(2\alpha H_0)} \times
              \hspace{1. cm}  \nonumber \\
             \exp \bigl[-\beta (2m+1)\alpha (q)\bigr] \ \sqrt{\alpha (q)}
                         \   \psi (q)   \hspace{1. cm}
                         \label{(89)}
                            \end{eqnarray}

\noindent
The factor $ 2, $ in  $ \widehat{(2\alpha H_0)},$  in the above  expression
occurs  because (\ref{(27b)}) was defined for $  t_2- t_1=\epsilon_1+\epsilon_2
= 2\epsilon. $ Thus we have,
             \begin{equation}
              \int {\cal K}[H_0,1,\alpha](qt,q_00) \psi (q_0)d^n q_0
              = \psi (q) - \lim_{N \rightarrow \infty} (\frac{4itN}{\hbar} )
              \sum^{N-1}_{m=1}\widehat{I_m} \psi (q)  \label{(90)}
              \end{equation}
\noindent
where the operator $\widehat{I_m} $ is
\vspace{5 mm}
           \begin{eqnarray}
          \lefteqn{ \widehat{I_m}  \equiv \int^{\infty}_{0} \beta  d\beta
            \Bigl\{ \alpha (q) \exp [-\beta  (2N-2m-1) \alpha (q) ] \Bigr\}
            \times \hspace{2.5 cm} }
             \nonumber \\
          &&
            \hspace{1. in}  \left(   \frac{1}{\sqrt{\alpha}} \
            \widehat{(\alpha H_0 )}
             \ \frac{1}{\sqrt{\alpha}}
               \right) \Bigl\{ \alpha (q) \exp[-\beta (2m+1)\alpha (q) ]
               \Bigr\}
                      \hspace{ 0.5 in}             \label{(91)}
                          \end{eqnarray}

To do the summation over m, we first pull $\left\{ \beta \ exp[-\beta (2m+1)
\alpha (q) ] \right\} $ across $ \alpha ^{-1/2} \ \widehat{(\alpha H_0 ) } \
\alpha ^{-1/2}  $ to the left, perform  $ \beta  $ integration  and  then take
the limit $ N \rightarrow \infty. $ Commuting the factor $ { \beta\exp[-\beta
(2m+1) \alpha (q) ]} $ across the operator $ {\alpha}^{-1/2}
\ \widehat{(\alpha H_0 )} \ {\alpha}^{-1/2}  $ to the left gives
                  \begin{equation}
                  I_m =\int^{\infty}_{0} d\beta  \beta ^2 \exp [-2\beta
                  N\alpha (q) ]
                  \left( \frac{1}{\sqrt{\alpha}} \widehat{(\alpha
                  H_0)}\frac{1}{\sqrt{\alpha}}
                   \right)  + \cdots
                          \label{(92)}
                  \end{equation}

             \noindent
where  $ + \cdots $ denote the terms which do not contribute when
(\ref{(92)})  is substituted in (\ref{(90)}) and limit $  N \rightarrow
\infty $ is  taken.  Therefore,  the terms of order $ t  $ in (\ref{(90)})
take the form
             \begin{eqnarray}
             \lefteqn{
              \lim_{N \rightarrow \infty} \sum^{N-1}_{m=0} (4N)
              \widehat{I_m}
               \psi \hspace{3. in} }  \nonumber
              \\
              && = \lim_{N \rightarrow \infty} (4N) \sum_{m=0}^{N-1}
              \int^{\infty}_{0}
             \beta  d\beta  \alpha (q)  \exp[-\alpha (2N-2m-1)\alpha ]
              \times  \nonumber
              \\
              && \hspace{1.5 in}  \bigl( \alpha ^{-1/2} \ \widehat{( \alpha
              H_0 )} \
              \alpha ^{-1/2}
             \bigr)  \alpha \exp[-\alpha (2m+1)\alpha ] \psi \nonumber
            \\
            && =  \lim_{N \rightarrow \infty} (4N)\sum^{N-1}_{m=0}  \int
            ^{\infty}_0 \beta
             d\beta \alpha ^2 \exp \bigl[ -2\beta  N\alpha (q)  \bigr]
             \left( \alpha
             ^{-1/2} \widehat{\alpha H_0} \alpha ^{-1/2}\psi  \right) +
             \cdots
             \label{(75a)}
             \\
              && = \bigl( \alpha ^{-1/2} \widehat{ \alpha H_0 } \alpha ^{-
              1/2}
                   \bigr)
          \psi   \hspace{3.5 in}
                               \label{(93)}
            \end{eqnarray}

\noindent
where we have used (\ref{(92)}) and
             \begin{equation}
                 \widehat{(\alpha H_0)}  = \frac{\partial}{\partial q^i}
                 \bigl( \alpha
                  g^{ij} \frac{\partial}{\partial q^j}\bigr)
                  \label{(94)}
       \end{equation}

\noindent
Thus we have the result
             \begin{eqnarray}
           \lefteqn{  \int {\cal K}[H_0,1,\alpha](qt,q_00)\psi (q_0)d^mq_0 }
            \nonumber \\ &&  = \psi (q) -  \frac{it}{\hbar}
           \bigl\{\alpha ^{-1/2}\widehat{(\alpha  H_0)} \alpha ^{-1/2}
           \bigr\}\psi (q) + O(t^2) \hspace{1.7 in}
                \label{(95)}
             \end{eqnarray}

\noindent
This above analysis could have been done with a potential term $ V $ added
to the Hamiltonian H , without altering anything.

The proof of the proposition 3.1 is completed by the  proving  that the
terms omitted in (\ref{(75a)}) do  not  contribute  to  the  final
answer (\ref{(93)}). For this purpose we shall first calculate  the
effect of commuting the factor  $ \exp[-\beta (2m+1)\alpha ] $ to the  left
across $\widehat{\alpha H} $ in (\ref{(91)}). We note that
                  \begin{eqnarray}
             \widehat{(\alpha H_0)}  \alpha  \exp[-\beta (2m+1)\alpha (q) ]
                  = \beta  \exp[-\beta (2m+1)\alpha (q) ] \times \hspace{1.0
in}
              \cr
                \left[ \widehat{\alpha H_0}
                +\{1-\beta (2m+1)\alpha \} 2g^{ij}\alpha _{ij} + \{1-\beta
                (2m+1)\alpha \}
                   (\alpha g^{ij} ),_i ( \alpha ,_j  /\alpha ) \right.
                \cr \left.
               + \{\beta ^2(2m+1)^2\alpha  - (2m+1)2\beta \}  g^{ij}\alpha
               ,_i \alpha,_j
               + \{1-\beta (2m+1)\alpha \}g^{ij}\alpha ,_{ij} \right]
               \label{(96)}
                 \end{eqnarray}

          \noindent
Substituting (\ref{(96)}) in (\ref{(91)}), the relevant integrals for us are
             \begin{eqnarray}
             \lefteqn{ \lim_{N \rightarrow \infty}(4N)  \sum^{N-1}_{m=0}
             \int^{\infty}_0
             \beta d\beta  \alpha ^2 \exp[-2N\beta \alpha ] } \nonumber
             \\
            &&  =  \lim_{n \rightarrow
             \infty}(4N)N\alpha ^2  \frac{1}{(2N\alpha )^2}  \nonumber
              \\
              &&  = 1  \hspace{4.2 in} \label{(97)}
                                                 \end{eqnarray}
            \begin{eqnarray}
            \lefteqn{ \lim_{n \rightarrow \infty}(4N)  \sum^{N-1}_{m=0}
            \int_0^{\infty}
            \beta d\beta  \alpha ^2 \exp[-2N\beta \alpha ](1-\beta
            (2m+1)\alpha )
            }  \nonumber \\
            &&  =1 -\lim_{n \rightarrow \infty}4N(N^2+O(N))
                        \frac{2\alpha ^3}{(2N\alpha )^3}  \nonumber
          \\ && =  0   \hspace{4.2 in}                    \label{(98)}
                       \end{eqnarray}
           \begin{eqnarray}
            \lefteqn{ \lim_{n \rightarrow \infty}(4N) \sum^{N-1}_{m=0}
            \int^{\infty}_0 \beta
             d\beta  \alpha ^2 \exp[-2N\beta \alpha ] \bigl\{\beta ^2
             (2m+1)^2 \alpha
             -2\beta (2m+1) \Bigr\}
             }   \nonumber  \\
            &&   =   4N\alpha ^3 \Bigl(\frac{4}{3} N^3 + O(N^2) \Bigr)
               \frac{6}{(2N\alpha)^4}  - 4N\alpha ^2 2(N^2+O(N))
               \frac{2}{(2N\alpha)^3}
               \nonumber
               \\ && = 0
                     \label{(99)}
               \end{eqnarray}

          \noindent
\newpage

Thus  all the terms in (\ref{(96)}), except the  first  one, vanish when the
beta integral is done and as $ N \rightarrow \infty $ is taken. It is proved
that  the  terms not written  explicitly  in (\ref{(75a)}) do not contribute
to the final result (\ref{(93)}).

The equation (\ref{(95)}) gives the desired result for $\rho=1 $. The
proof for  a general $\rho$ follows if use is made of the fact that the
$\rho$ dependence of the ${\cal K}[H,\rho,\alpha]$ is simple and is given by
         \begin{equation}
         {\cal K}[H,\rho,\alpha] = \frac{1}{\sqrt{\rho(q)\rho(q_0)}}
                                     {\cal K}[H,1, \alpha]
         \end{equation}

When  the  hamiltonians of interest is quadratic in  momenta an alternative,
but  considerably  more  complicated,  proof  of the above propositon can be
given  by  first  converting  the  STP into an equivalent Lagrangian form by
doing  the momentum integration in the STP and converting it to the standard
mid  point  form using the McLaughlin Schulmabn trick. Next  N- fold product
of  this  form  of STP, with  intermediate $N-1$ fold integrations over $ q$
variables,  is  substituted for the HPI1 appearing in the right hand side of
(\ref{(72)}). Repeated use of McLauglin Schulman after carefully identifying
the  important terms for small $t$ and performing all the integrations again
once leads to the same result in the limit $ N \rightarrow \infty $. For more
details on this method we refer the reader to ref 38.

It must be  emphasized  that  though  $ t $ can be taken to be small for the
present purpose,  we cannot take $\sigma$ small  and replace HPI1 in
(\ref{(72)}) by a {\it single } STP. A useful corollary of the proposition is
 obtained by taking $ \alpha = \rho=\sqrt{g}$ leading to the result

{\it Proposition 3.2 :} For the hamiltonian function $  H_0 = \frac{1}{2m}
 g^{ij} p_i p_j, $ the HPI2 $ {\cal K}[H_0, \sqrt{g},\sqrt{g}], $ has the
 normalization
         \begin{equation}
\lim _{t \rightarrow t_0} {\cal K}[H_0, \sqrt{g},\sqrt{g}](qt;q_0t_0)
  = g^{-1/2}  \delta(q-q_0)
                                         \end{equation}
\noindent
and satisfies the Schrodinger equation
         \begin{eqnarray}
            i\hbar \frac{\partial \psi}{\partial t}
              & = & g^{-1/2}\  \widehat{(\sqrt{g}H_0)} \ )\psi
                                        \nonumber    \\
             i\hbar \frac{\partial \psi}{\partial t}
             &  =  &- \frac{\hbar ^2}{2m}  g^{-1/2}
              \frac{\partial}{\partial q^i} \left( g^{1/2} g ^{ij}
              \frac{\partial }{\partial q^j}  \psi \right)
                \label{(47)}
                                                  \end{eqnarray}
          \noindent
For later use we note that the HPI2 $ {\cal K}[H,\sqrt{g},\sqrt{g}] $ can
also be written as
\vspace{1 in}
                  \begin{eqnarray}
          {\cal K}[H, \sqrt{g},\sqrt{g} ](qt,q_00) \hspace{3.5 in} \cr
           = \int_{- \infty}^{\infty} \frac{dE}{2 \pi \hbar} \exp(-iEt/\hbar)
           \int_0^{\infty} d\sigma  K[\alpha (H-E),1](q\sigma ,q_0 0)
           \hspace{.5 in}
               \label{(556)}
             \end{eqnarray}

\noindent
The  path integral (\ref{(556)} was at first introduced in paper II where it
was  proved,  using  a  different  approach,  that  it satisfies the correct
Schrodinger equation in arbitrary co-ordinates.

{\it  Proposition 3.3 : } The HPI2 $ {\cal K}[H,\rho,\alpha] $  is related
to HPI1 without scaling by means of formula
          \begin{equation}
            {\cal K}[H,\rho,\alpha ] = K[H-U(\alpha), \rho]
            \end{equation}
The form of $  U(\alpha) $ depends on the Hamltonian function
$ H(q,p).$ For $ H(q,p) $ given by (\ref{(40)}) the function $ U(\alpha) $
is defined by (\ref{(44)}). The above result is proved by showing that both
sides have the same  normalization and satisfy the same Schrodinger equation.

          {\it Proposition 3.4 :} For a trivial scaling when the scaling
          function
          $\alpha(q) =  {\rm \ constant,\ say, \ } c $  the HPI2 $ {\cal
          K}[H,\rho,c] $ is equal to the HPI1 $ K[H,\rho] $ without
          scaling.

          \begin{equation}
          {\cal K}[ H,\rho,c] = K[ H,\rho]
          \end{equation}

          \noindent
          This equation with  $ \alpha = $ constant is easily proved by
          following the  proof of the proposition 3.1 upto equation
          (\ref{(84)}) and   noticing that all the
          terms involving the derivatives of $ \alpha(q) $ vanish and thus
          (\ref{(84)}) leads to the discrete form (\ref{(28)}).

 \vspace{5 mm}
\noindent
              {\bf 4. An example of quantization in arbitrary coordinates}

 \vspace{5 mm}

The  equation (\ref{(47)}) is recognised as the correct Schrodinger equation
in  arbitrary  co-ordinates.  This  makes  HPI2  $  {\cal  K}  $  useful for
quantization  in arbitrary coordinates. In some special cases   $  {\cal  K}
$    itself  is  equal  to the correct propagator in arbitrary   cordinates.
In   general   situations,  more frequently  encountered, the propagator can
be easily written in terms of the HPI2 $  {\cal  K}  $   defined  above.  As
an  example  of quantization in arbitrary  coordinates  and a simple example
of exact path integration we set up the  path  integral  quantization  of  $
1/r   $  potential  in parabolic  coordinates  in  two dimensions obtain the
exact answer for the Green function.

The   classical   Hamiltonian   for  the  two dimensional Coulomb problem is
                   \begin{equation}
                                      H  =\frac{\vec{p}^2 }{2m} -
                                      \frac{e^2}{r}
                       \label{(187)}
                         \end{equation}
               In the two dimensional parabolic coordinates, $ u_1 $
               and  $ u_2, $  defined by
               \begin{eqnarray}
                             x = u^2_1 - u^2_2 & , &  y = 2 u_1 u_2
                             \label{(188)} \\
               dx dy  =  4 \vec{u}^2 du_1 du_2  & , &  g^{1/2} = 4u^2 =
               u^2_1+u^2_2
                                          \label{(6.5)}
                                          \end{eqnarray}
              the classical hamiltonian takes the form
              \begin{equation}
                        H =  \frac{ \vec{p} ^2 _u }{8m u^2} -
                        \frac{e^2}{u^2}               \label{(6.4)}
                                          \end{equation}

                Quantization  in parabolic coordinates $ u_1,u_2 $ will
                proceed  via
              path  integral  HPI2 with hamiltonian $ H, $ scaling function
              $ \alpha \equiv  4  \vec{u}^2.  $  Therefore  we  consider
              the hamiltonian path integral of the second kind
             \begin{equation}
             {\cal K} [H ,4u^2,4u^2](\vec{u}t; \vec{u}_0 0) = \int
             \frac{dE}{(2
             \pi \hbar)} \exp(-
             iEt/\hbar ) \int_{0}^{\infty} d\sigma  K[ H_E, 1](\vec{u}
             \sigma
             ;\vec{u}_0 0)  \label{(6.7)}    \end{equation}

\noindent
              with
             \begin{equation}
                        H_E  =  4u^2 (H  - E)   = \left(\frac{\vec{p}^{
                        ~2}_u}{2m}- 4e^2 - 4E u^{2} \right)
\label{(6.8)}
\end{equation}

Notice  that  apart from an  additive  constant, $ H_E $ is  just  the
harmonic  oscillator hamiltonian in two dimensions. Thus we  see  that
the   HPI1   in (\ref{(6.7)})  is  related  to  the  propagator  for  the two
dimensional  oscillator  and the relation is given by

                        \begin{equation}
                           K[H_E,\rho =1]( \vec{u} \sigma ;\vec{u}_0 0) =
                           \exp(i4e^2\sigma
             / \hbar ) {\bf k}^{osc}\langle u_1\sigma \vert u_{10}0 \rangle
             {\bf
             k}^{osc} \langle u_2 \sigma \vert u_{20} 0 \rangle
             \label{(6.9)}
                                         \end{equation}

             \noindent
with  $  {\bf k}^{osc} $ is the one dimensional oscillator propagator for
mass  $ \mu=m $  and  frequency  $  \omega^2  =  -8E/M.  $ Though  HPI2 of
(\ref{(6.7)}) satisfies the correct Schrodinger equation, it is still
not  equal  to  the desired propagator. This is because at $ t=0  $ it
reduces  to  $  g^{-1/2} \delta (\vec{u}-\vec{u}_0)  $  where as the correct
propagator at $ t=0 $ becomes

         \begin{equation}
                         \delta (\vec{r} -\vec{r}_0)  =  g^{-1/2} \left[
                         \delta
                         (\vec{u}-\vec{u}_0) +\delta  (\vec{u}+\vec{u}_0)
                         \right]
                         \label{(6.11)}
             \end{equation}

It  is  very  easy  to  take  care  of this difference and the correct
propagator can be written down as
             \begin{equation}
              \langle \vec{r}t \vert\vec{r}_0 0 \rangle =  \bigl[ {\cal K}
              [H_0
             ,4u^2, 4u^2](\vec{u} t; \vec{u}_0 0)  +  {\cal K} [H_0 ,4u^2,
             4u^2](\vec{u} t;\vec{u}_0 0) \bigr]
       \label{(6.12)}
\end{equation}
\newpage

 Thus  the  desired  quantum  mechanical propagator $  \langle
 \vec{r}t \vert\vec{r}_0  0  \rangle   $    is obtained from
 (\ref{(6.12)}) and using the expressions (\ref{(6.7)}) and (\ref{(6.9)}).
 Writing the  answer in terms of the energy dependent Green function we get
             \begin{eqnarray}
             {\bf G} (\vec{r},\vec{r}_0\vert E)
             = \int_{0}^{\infty} d\sigma \exp(4ie^2\sigma /\hbar )\bigl[
             {\bf
             k}_1^{osc}\langle u_1
              \sigma \vert u_{10}0 \rangle {\bf k}_1^{osc} \langle u_2
             \sigma \vert u_{20}0 \rangle    \cr
              +   {\bf k}_1^{osc}\langle u_1 \sigma \vert -u_{ 10}0\rangle
              {\bf
              k}_1^{osc}\langle u_2 \sigma \vert -u_2 0\rangle \bigr]
              \label{(6.13)}
             \end{eqnarray}

          \noindent
After the integration  variable   $  \sigma  $ is replaced by  $ \tau =
4\sigma,$  this  result  agrees   with  the result of ref. 13.

\vspace{5 mm}
\noindent
          {\bf    5.  Concluding remarks}

\vspace{5 mm}

   In this paper we have discussed properties of two types of path integrals
   HPI1  and HPI2. The path integral HPI2 intoduced here is a generalization
   of  HPI1  introduced  in  paper  I. An importatnt result obtained in this
   paper is the scaling formula of proposition 3.3. We have not followed the
   approaches  existing   in   the literature to  derive the scaling formula
   or  a  similar result in the path integral or the operator formalism. The
   propagators  appearing  in the two sides of this formula were constructed
   using  path  integration with the same prescription for STP. We have then
   derived  the effective potential that the scaling formula  introduces  in
   the  classical  Lagrangian  or  Hamiltonians. The method  used by us is a
   direct  approach starting from the definintions used and can also be used
   for Lagrangian path integrals.

   Our  use  of the path integral with scaling also differs  from the use in
   the  existing  literature  where  local  scaling  is  merely  one  of the
   techniques  to   be   used  in exact path integration. For us  the  local
   scaling  is  an  essential   ingredient   for  quantization  by  means of
   hamiltonian   path   integration  in  arbitrary coordinates. Use of local
   scaling  in  the   hamiltonian   path   integral   scheme  enables  us to
   formulate path integral  quantization  scheme  in  terms of the classical
   hamiltonian  only,  that  no  $ O(\hbar^2) $ terms are needed is a unique
   feature  of  this  scheme.  At  present  the origin of nontrivial scaling
   necessary  in  the hamiltonian formalism for non-cartesian coordinates is
   not understood from any deeper physical or mathematical reason.

   We   believe that the Hamiltonian path integral formalism  can be used to
   discuss everything, and at the same level of  rigor, as in the Lagrangian
   approach.     To     support    this   claim   we should demonstrate  the
   applicability   of   the  Hamiltonian method  to  exact path  integration
   of   various   potential  problems.   In  the  existing literature  there
   are    other   techniques,  notably,  the  trick  of adding  new  degrees
   of  freedom,  which also play an important role in  exact  path  integral
   evaluations.  However, these   are   best  discussed  for  each  specific
   case  of   exact  path  integration. In this  paper  we  have  restricted
   ourselves  to  the  local  rescaling  of time  only  because  it  is  the
   general   idea   common   to  all  such  calculations.  A  discussion  of
   other  techniques  together  with  exact  path   integration of potential
   problems will be taken up in a separate publication$^{38}.$

   It  is  not  just  that  it  may  be  possible  to  do everything  in the
   hamiltonian   approach   that   can   be   done  in   existing Lagrangian
   formalism;   the   Hamiltonian   approach  offers distinct advantages  in
   certain   problems.   As   is  well known  the  constraint  analysis  and
   unitarity  are  most  transparent in phase space path integral formalism,
   and   it  should   be possible  to give a time slicing  definition of the
   formal  path  integrals  of Faddeev$^{39}$ and of Senjanovic $^{40}$  for
   the constrained systems.

   To   elaborate  on  the  Hamiltonian  path integral with local scaling of
   time   and  show  its  efficacy in handling quantization in arbitrary co-
   ordinates   has   been   the   main   objective  of  this  paper.  In the
   literature   the   hamiltonian  path integral scheme have not received as
   much   attention   as   the   Lagrangian  path  integral  approach, which
   incidentally,  is  reflected  in  the  sketchy  treatment  given  to  the
   canonical    path  integral in books such as by Schulman$^{7}.$ We fondly
   hope   that,     with    our   presentation  here,  the  impression  that
   the Hamiltonian   path integral is meant only for a formal discussion wil
   be removed.

   {\it Acknowledgment:} One of the  authors (A.K.K.)  thanks  Prof. H.S.Mani
     for hospitality extended to him at the Mehta Research Institute.
          \pagebreak

          \noindent
   {\bf REFERENCES }
   \vspace{1 cm}
  \begin{enumerate}
  {\item  P.A.M. Dirac,{\it  Principles  of  Quantum Mechanics,}
          Calrendon, Oxford, 1958; R.P. Feynman,
          Revs. Mod. Phys.  {\bf 20} 365  (1948);
          R.P. Feynman and  A.R.  Hibbs, {\it Quantum  Mechanics  and  Path
          Integral,} McGraw Hill, New York, 1965. }
 {\item  S.  Albeverio  and  R.  Hoegh-Krohn,
        {it Mathematical  Theory  of Feynman Integrals,}
        Lecture Notes in Maths. vol. {\bf 523}, Springer, Berlin, 1976;
        A. Truman, J. Math. Phys. {\bf 17}, 1852 (1976);
        ibid, {\bf 18}, 2308 (1977); ibid {\bf 19}, 1742 (1978);
        C. Dewitt-Morette,  A. Maheswari and B. Nelson,
        Phys.  Reports  {\bf 50}, 255 (1979) ;
        I. Daubechis and J.R. Klauder, J. Math. Phys. {\bf 26}, 2239 (1985). }
{\item M.S. Marinov, Phys. Reports {\bf 60}, 1 (1980) }
{\item  D.C. Khandekar and S.V. Lawande, Phys. Reports {\bf 137}, 115 (1986).}
{\item A.S.Arthurs (ed.),{\it Functional Integration and its Applications,}
       Proceedings of the International  Conference  at London 1974,
       Oxford University press 1975. }
{\item G.J. Papadopoulos and J.T. Devrese (eds.), {\it Path Integrals  and
       Their Applications to Quantum, Statistical and Solid State Physics,}
       Plenum, New York, 1978. }
{\item L.S.   Schulman,{\it   Techniques and Applications of Path Integration,}
       John Wiley, New York 1981. }
 {\item M.C. Gutzwiller, A.  Inomata,  J.R.  Klauder, and L. Streit (eds.),
       {\it Path Integrals from  meV  To MeV,}  World  Scientific, Singapore,
       1986. }
{\item S. Ludqvist et al., (eds.),{\it Path Integral Method and Applications,}
       Proceedings of Adriatico  Conference  on  Path Integrals,  Trieste,
       Italy,   1-4   September   1987,   World Scientific,
       Singapore, 1988. }
{\item For early evaluations of non-gaussian path Integrals see,
       D.C. Khandekar, and S.V. Lawande, J. Phys. {\bf A5} 812 (1972);
       {\bf A5}, L57 1972;
       A. Maheswari, J. Phys {\bf A8}, 1019 (1975); see also ref. 11.}
{\item D. Peak and A. Inomata, J. Math. Phys. {\bf 10}, 1422(1969). }
{\item I.H.  Duru  and  H.  Kleinert,  Phys.  Lett.{\bf 84B}, 185 (1979);
       Fortschr. der Phys. {\bf 30}, 401 (1982). }
{\item P. Kustanheimo and E. Stiefel, J. Reine Angew Math. {\bf 218},
       204 (1965). }
{\item  R. Ho and A. Inomata, Phys. Rev. Lett. {\bf 48}, 231 (1982). }
{\item  A. Inomata, Phys. Lett. {\bf 87A}, 387 (1982).}
{\item  A. Inomata, Phys. Lett.{\bf 101A} , 253 (1984);
        F.  Steiner,  Phys. Lett. {\bf 106A}, 363 (1984). }
{\item  H. Kleinert, Phys. Lett. {\bf 116A}, 201 (1986). }
{\item  I. Sokmen, Phys. Lett. {\bf 132A}, 65 (1988).}
{\item  H. Durr and A. Inomata, J. Math. Phys. {\bf 26}, 2231 (1985). }
{\item  P. Y. Cai, A. Inomata, and R. Wilson, Phys. Lett.{\bf 99A}, 117 (1983)
     }
{\item  A. Inomata and M. Kayed, J. Phys. {\bf A18}, L235 (1985).}
{\item  A. Inomata and M. Kayed, Phys. Lett. {\bf 108A}, 117 (1983).}
{\item  M.V. Carpio and A. Inomata, see in Ref. 8.}
{\item  J.M. Cai, P.Y. Cai, and A. Inomata, Phys. Rev. {\bf A34}, 4621 (1986) }
{\item  S.V. Lawande and K.V. Bhagat, Phys. Lett. {\bf 131A}, 8 (1988);
        D. Bauch, Nuovo Cim. {\bf 85B}, 118 (1985). }
{\item  W. Janke and H. Kleinert, Lett. Nouvo Cim.{\bf 25} , 297 (199);
        I. Sokmen, Phys. Lett {\bf 106A}, 212  (1984). }
{\item  N.K.Pak  and I. Sokmen, Phys. Lett {\bf 103A}, 298 (1984);
         Phys. Rev. A30  1629 (1984);
         I.H. Duru, Phys. Lett. {\bf 112A}, 421 (1985);
         I. Sokmen, Phys. Lett. {\bf 115A}, 249 (1986);
         S. Erkoc and R. Sever, Phys. Rev. {\bf D30}, 2117 (1984);
         I. Sokmen, Phys. Lett. {\bf 115A}, 6 (1986) }
 {\item  A.K. Kapoor, Phys. Rev. {\bf D29}, 2339 (1984). }
 {\item  A.K. Kapoor, Phys. Rev. {\bf D30}, 1750 (1984).}

 {\item  N. Pak and I. Sokmen, Phys. Rev.{\bf A30}, 1629 (1984);
         J.M. Cai, P.Y. Cai, and A. Inomata, Phys. Rev. {\bf A34}, 4621 (1986)
         and references therein;
         Alice Young and Cecile Dewitt-Morette, Ann. Phys. {\bf 169}, 140
         (1986).}
 {\item  W. Pauli, {\it Selected Topics in field Quantization,}
          MIT  Press, Cambridge, Mass, 1973. }
 {\item  C.M. DeWitt, Phys. Rev.{\bf 81}, 848 (1951).}
 {\item  J.H. Van Vleck, Proc. Natl. Acad. Sci. USA {\bf 14}, 178 (1928).}
 {\item  J.S. Dowker, J. Phys.{\bf A3 }, 451 (1970).}
 {\item  B.S. DeWitt, Rev. Mod. Phys. {\bf 29}, 377 (1957).}
 {\item  D.W. McLaughlin and L.S. Schulman J. Math. Phys. {\bf 12}, 2520
          (1971) }
 {\item  A.K.Kapoor and Pankaj Sharan, Hyderabad University Preprint HUTP-86/5}
 {\item  A.K.Kapoor and Pankaj Sharan, {\it Hamiltonian path integral
         quantization in arbitrary coordinates and exact path  integration, }
         Mehta Research Institute Preprint MRI-PHY/18/94 . }

 {\item  L.D. Fddeev, Theor. Math. Phys. {\bf 1}, 1 (1969).}
 {\item  P.Senjanovic, Ann. Phys. (N.Y.) {\bf 100}, 227 (1976).}

 \end{enumerate}

\end{document}